\documentclass[letter,twocolumn]{jpsj3} %% two-column layout
\usepackage{mathrsfs} %\mathscr 
\usepackage{amsmath,amssymb}
\def\Norm{\nu}

\def\e#1{\,{\rm e}#1}

\def\vec#1{{\overrightarrow{#1}}}
\def\Tr{\mathop{{\rm Tr}}}

\def\and{\,\,\&\,}

\def\ket#1{|#1\rangle }
\def\bra#1{\langle #1|}
\def\braket#1#2{\langle #1|#2\rangle}

\def\i{{\bf i}}

\def\skipm#1{}

\def\bfsigma{\boldsymbol{\sigma}}
\def\bfalpha{\mathbf{\alpha}}

\def\section#1{}
\def\footnote#1{{\tiny #1}}
\def\vec#1{{\mathbf{#1}}}
\def\i{i}

\title{Uniform Matrix Product State in the Thermodynamic Limit}
\author{Hiroshi \textsc{Ueda},$^1$\thanks{E-mail address: ueda@aquarius.mp.es.osaka-u.ac.jp} Isao \textsc{Maruyama}$^1$\thanks{E-mail address: maru@mp.es.osaka-u.ac.jp} and Kouichi \textsc{Okunishi}$^2$\thanks{E-mail address: okunishi@phys.sc.niigata-u.ac.jp} }

\inst{$^1$Department of Material Engineering Science, Graduate School of Engineering Science, Osaka University, 
Toyonaka, Osaka 560-8531, Japan \\
$^2$Department of Physics, Faculty of Science, Niigata University, 
Niigata 950-2181, Japan \\
}
\abst{
We study a uniform matrix product state as a variational state for classical and quantum spin chains in the thermodynamic limit.
Under a careful treatment of the translational symmetry, eigen values of the transfer matrix defined in the calculation of expectation values can reflect the periodicity of the ground state and indicate optimum periodicity of the matrix product state.
We discuss the relation between the periodicity and accuracy of magnetization curves.
This approach is free from the error due to finite system size, which works well especially for the magnetic plateau problem.

}
\kword{matrix product state, iTEBD, DMRG, magnetization plateau, anti-ferromagnetic spin chian, variational method}

\begin{document}
\maketitle

\section{Introduction}
A recent trend of numerical calculations of quantum mechanics is a construction of new variational states as extensions of the matrix product state (MPS), such as the tensor product state (TPS),~\cite{Niggemann:ZPB104-EPJB13, Delgado:PRB64}
the projected entangled pair state (PEPS),~\cite{Verstraete:PRL96} the tree tensor network state (TTNS),~\cite{Shi:PRA74, 1006.3095} and the multi-scale entanglement renormalization ansatz (MERA) state~\cite{Vidal:PRL101}.
The most driving force of this trend is the success of the density matrix renormalization group (DMRG) method,~
\cite{White:PRL69-PRB48, Peschel:Springer, RevModPhys.77.259, Ostlund:PRL75, Rommer:PRB55} 
which is a powerful numerical method for one-dimensional strongly correlated systems.
The MPS is not only used as a variational state~\cite{baxter:650} but also the exact ground state of the valence-bond-solid (VBS) state in the Affleck-Kennedy-Lieb-Tasaki (AKLT) model~\cite{Kluemper:JPA.24}, and the exact eigen states of the Bethe ansatz.~\cite{Alcaraz:JPA.37-JPA.39, JPA.43.175003}

\section{Introduction: Uniform Matrix}
The extensions of the MPS in the thermodynamic limit have been widely studied, where we expect that the MPS has uniform, i.e., site independent, matrices for the uniform Hamiltonian, for example, in the infinite time-evolving block decimation (iTEBD)~\cite{PhysRevLett.98.070201}, the infinite PEPS (iPEPS)~\cite{Jordan:PRL101}, and scale invariant MERA.~\cite{Robert:PRA79}
Such a uniform matrix is obtained as a fixed point of the DMRG.~\cite{Ostlund:PRL75, Rommer:PRB55} 
As a merit of the uniform MPS, we can handle the infinite uniform system directly, that is, we can construct variational states with finite dimensional matrices in spite of the infinite dimensional Hilbert space of the infinite system.
In some methods, the matrices are optimized by the iterative projection with using Suzuki-Trotter checker-board decomposition\cite{PhysRevLett.98.070201,PhysRevLett.101.090603,Jordan:PRL101,PhysRevB.81.174411}.

\section{Introduction: plateau}
A natural question for the uniform MPS is how to deal with the spontaneous symmetry breaking of the translational symmetry,
where a ground state has long periodicity.
One of interesting examples is the magnetization plateau, which emerges in some frustrated systems and has been widely researched, because we can expect a new quantum phase in a new plateau. 
For example, in the $S=1/2$ antiferromagnetic zigzag chain, the novel even-odd effect of the 1/3 plateau has been studied by using the DMRG\cite{Okunishi:PRB68}.
In the calculation of the magnetization curve, the finite size effect causes the small steps in the magnetization.
On the contrary, the uniform MPS in the thermodynamic limit gives a smooth magnetization curve by definition, 
which is an advantage of the uniform MPS.

\section{In this letter}
In this letter, we study the uniform MPS with arbitrary periodicity and a boundary matrix for magnetic plateaus.
To treat the translational symmetry explicitly,
we do not use the checker-board decomposition, which breaks the primitive periodicity, i.e., the one-site translational symmetry.
When we consider a linear combination of $q$-fold degenerated ground states with $q$-site periodicity,
the periodicity of the MPS is determined by the boundary matrix. 
The boundary matrix is important also in discussions of the fixed point of the DMRG~\cite{Ostlund:PRL75, Rommer:PRB55} and in the continuous MPS~\cite{JPSJ.79.073002}.
The MPS used in this letter
has matrices  generalized toward Baxter's interaction-round-a-face(IRF)-type MPS,~\cite{Baxter:book} which has an advantage in the case of dimerization. 
Using the IRF-type MPS as a generalization of the usual MPS, we demonstrate the performance to calculate the magnetization curve in quantum spin chains. 
From the numerical point of view, it is important to consider the optimum periodicity of the MPS in order to obtain the best variational ground state.

\section{MPS and IRF-Type MPS}
Let us start from the definition of a finite $N$-site variational state $\ket{\Psi}$ with the boundary matrix $A_0$.
The total Hilbert space for a spin $S$ system is spanned by $\ket{\bfsigma}=\prod_{i=1}^N \ket{\sigma_i}$ with $(2S+1)$-dimensional local basis $\ket{\sigma_i}$, where $\bfsigma=\{\sigma_1,\ldots,\sigma_N\}$.
Many theoretical studies have tried to reduce computational memory by using tensors $A$ with artificial local bases $\alpha_i$ instead of
$(2S+1)^N$ dimensional vector $\Psi(\bfsigma) = \braket{\bfsigma}{\Psi}$.
Generally, the number of tensors $A$ and local bases $\alpha_i$ depend on the coordination number and the network of tensors. 
In one dimensional case, a simple MPS without $A_0$ is defined as
\begin{math}
  \Psi(\bfsigma)  = \sum_\bfalpha \prod_{i=1}^N A^{\sigma_i}_{i;\alpha_i,\alpha_{i+1}}
,
\end{math}
where $\bfalpha=\{\alpha_1,\ldots,\alpha_N\}$, and $\alpha_{N+1}=\alpha_{1}$.
We suppose the dimension of each $\alpha_i$ is finite $m$.
At $m=(2S+1)^N$ any $\Psi$ will be expressed.
There are other expressions such as $\Psi(\bfsigma) = \Tr \left[\prod_{i=1}^N A_i^{\sigma_i} \right]$ with $(A_i^{\sigma_i})_{\alpha,\alpha'} = A^{\sigma_i}_{i;\alpha,\alpha'}$, and $\ket{\Psi} = \Tr\left[\prod_{i=1}^N A_i \right]$ with $(A_i)_{\alpha,\alpha'} = \sum_{\sigma_i} A^{\sigma_i}_{i;\alpha,\alpha'} \ket{\sigma_i}$.

As a generalization of the usual MPS, we use a IRF-type MPS
\begin{eqnarray}
{\textstyle 
\Psi(\bfsigma)  = \sum_{\bfalpha} A^{\sigma_{N},\sigma_{1}}_{0;\alpha_{N},\alpha_{1}}
\prod_{i=1}^{N-1} A^{\sigma_i,\sigma_{i+1}}_{i;\alpha_i,\alpha_{i+1}}
},
\label{eq:def:Psi}
\end{eqnarray}
where $\bfalpha=\{\alpha_1,\ldots,\alpha_{N}\}$ are artificial local bases, and $A_0=A_N$ is a generalized boundary matrix.
We call it a IRF-type MPS in distinction from the usual MPS with the boundary matrix\cite{Ostlund:PRL75, Rommer:PRB55}, which is represented as the special case that $A_{i;\alpha\alpha'}^{\sigma\sigma'}=A_{i;\alpha\alpha'}^{\sigma}$ and $A_{0;\alpha\alpha'}^{\sigma\sigma'}=\sum_{\alpha''}A_{N;\alpha\alpha''}^{\sigma}\Omega_{\alpha''\alpha'}$.
The IRF-type MPS is an extension of the usual MPS as shown in Fig.~\ref{fig:1}, and can represent local entanglement very well as shown below.
\begin{figure}
  \centering
  \resizebox{8cm}{!}{\includegraphics{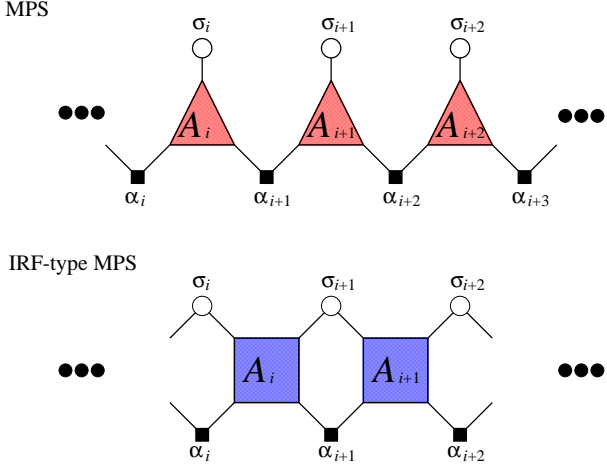}}\\
  \caption{Graphical representations of the usual MPS and the IRF-type MPS. 
Filling the small square means contraction with respect to the local artificial bases $\alpha_i$.}
  \label{fig:1}
\end{figure}

To demonstrate the ability of the MPS at small $m$ we consider 
the direct product state of the local singlets
which is the exact ground state of
the spin-1/2 dimerized Hamiltonian ${\cal H}_{\rm D}=\sum_i \vec{S}_{2i-1}\cdot \vec{S}_{2i}$.
Corresponding the two-site translational invariance of ${\cal H}_{\rm D}$, we suppose $A_{2i-1} = A_{1}$, and $A_{2i}= A_2$ for all $i$. 
The usual MPS at $m=1$ is just a direct product state of the local spins,
which is a classical state with no quantum entanglement.
At least, $m=2$ is required to represent the exact ground state for the usual MPS.~\cite{comment:MPS1}
The IRF-type MPS can represent the exact ground state even in the $m=1$ case.~\cite{comment:MPS2}
Here, $A_0$ is identical and is not important.

To show the importance of $A_0$,
let us consider the doubly degenerated classical N\'{e}el states expressed by the IRF-type MPS,
which is realized at $m=1$
by $A_{i}^{\uparrow\downarrow}= A_{i}^{\downarrow\uparrow}=1$, and $A_{i}^{\uparrow\uparrow}= A_{i}^{\downarrow\downarrow}=0$.
Then, we obtain
$\ket{\Psi}=A_{0;1,1}^{\downarrow\uparrow}\ket{\uparrow\downarrow\cdots}
+ A_{0;1,1}^{\uparrow\downarrow}\ket{\downarrow\uparrow\cdots}$
for even $N$.
We emphasize the boundary matrix $A_0^{\sigma\sigma'}$ determines the periodicity of $\ket{\Psi}$,
and $\ket{\Psi}$ has one-site translational invariance if $A_0^{\downarrow\uparrow}=A_0^{\uparrow\downarrow}$.

In general, there is the exact mapping from the IRF-type MPS with $m=m_0$ to the usual MPS with $m=(2S+1) m_0$ via the singular value decomposition (SVD). 
If we apply the SVD to IRF-type MPS as
$A^{\sigma\sigma'}_{i\alpha\alpha'} = \sum_{\beta}
\bra{\sigma\alpha}U_{i}\ket{\beta}\gamma_{i\beta}
\bra{\beta}V_i^\dagger \ket{\sigma'\alpha'}$
with singular values $\gamma_{i\beta}$, we obtain the usual MPS 
$A^{\sigma}_{i\beta\beta'}
=\sum_{\alpha}
\sqrt{\gamma_{i-1\beta}}
\bra{\beta}V_{i-1}^\dagger \ket{\sigma\alpha}
\bra{\sigma\alpha}U_{i}\ket{\beta'}\sqrt{\gamma_{i\beta'}}
$,
where the dimension of the index $\beta$ is the same as $\sigma\alpha$, that is, $(2S+1) m_0$.

\section{Normalization and Expectation Values}
Let us summarize the formulations of the IRF-type MPS with the finite period $p$ and the dimension $m$ to deal with the thermodynamic limit $N\rightarrow \infty$.
We suppose translational invariance with period $p$ as $A_i=A_{i + n p}$ for all non-negative integer $i$ and $n$.
The system size has the relation $N=N_1 p + N_0$ with the non-negative integer $N_1$ and $N_0 = N \mod p$.
We consider the mismatch $N_0$ generally, but the effect due to $N_0$ will be merged into the boundary effect due to $A_0$ as shown later.
For simplicity, we consider  $A_{0}=A_{p}$ and $N_0=0$ for a while.

For a finite $N$, the inner product is written by
\begin{math}
  {\braket{\Psi}{\Psi}}
  =
  \Tr\left[\prod_{i=1}^N T_i\right]
\end{math}
with a transfer matrix $T_i$ with matrix elements
\begin{eqnarray}
  (T_{i})_{\alpha\alpha''\sigma, \alpha'\alpha'''\sigma'} &=&   
  (A_{i;\alpha\alpha'}^{\sigma\sigma'})^*
  A_{i;\alpha''\alpha'''}^{\sigma\sigma'}
  .
\end{eqnarray}
Using the $p$ times periodicity for the $N=N_1 p$ system, we have
\begin{math}
  \braket{\Psi}{\Psi}
  =
  \Tr  \left[T^{N_1}\right]
\end{math}
with \begin{math}
  T = \prod_{i=1}^p T_i \;
  .
\end{math}

In the thermodynamic limit, only principal eigenvalues $\lambda_j$ of $T$ is important,
where ``principal'' means that $|\lambda_j|=\lambda$ is maximum.
Hereafter, we suppose $T$ is diagonalizable.
We denote principal eigenvalues as $\lambda_j=\lambda \e^{\i \theta_j}$ and corresponding left (and right) eigenvector as $\vec{u}_j$ ($\vec{v}_j^\dagger$) with the bi-orthogonal condition $\vec{v}_j^\dagger \vec{u}_{j'} = \delta_{jj'}$.
At large $N_1$, we have 
\begin{eqnarray}
{\textstyle 
  T^{N_1} \simeq \sum_j \vec{u}_j   \lambda^{N_1} \e^{\i N_1 \theta_j}
  \vec{v}_j^\dagger
} .
\end{eqnarray}
Hereafter, we renormalize $\tilde{A}_i=\lambda^{{1\over 2p}} A_i$ and rewrite it as $A_i$.
Then, we obtain
\begin{math}
  \lim_{N\rightarrow\infty }
  \braket{\Psi}{\Psi}
  =
  \sum_j \Norm_j
\end{math}
with normalization
\begin{math}
{\textstyle
  \Norm_j =
  \lim_{N \rightarrow\infty }
  \e^{\i \theta_j N_1 } 
}.
\end{math}
By definition, $N\rightarrow \infty$ means $ N_1 \rightarrow \infty$.
For a general phase $\theta$, we can adopt 
\begin{eqnarray}
  \lim_{N_1 \rightarrow\infty }  
  \e^{\i \theta N_1 }
  =
  \left\{
    \begin{array}{cc}
      1  &, \theta=0
      \\
      0 &, \theta\neq 0
    \end{array}
  \right. .
\end{eqnarray}
However, a normalization $\nu_j$ for a commensurate phase $\theta_j = 2\pi j/q$
is able to be non-zero if we consider the special case $N_1 \propto q$.
This means that emergence of the commensurate phase indicates $q p$-periodicity of the state.

To control the periodicity in the thermodynamic limit, 
we use the boundary operator $A_0$.
We can define the boundary transfer matrix
$T_0$ for $p+N_0$ sites
including $A_0$ and the mismatch $N_0$.
Then, the norm
is written as
\begin{math}
  \braket{\Psi}{\Psi}
  =
  \Tr  \left[T_0 T^{N_1-1}\right]
  =
  \sum_j \Norm_j
\end{math}
with
\begin{eqnarray}
{\textstyle
  \Norm_j =
  \lim_{N \rightarrow\infty }
  \e^{\i \theta_j (N_1 - 1)}  \vec{v}_j^\dagger
  T_0 \vec{u}_j
  \label{def:N_i}
},
\end{eqnarray}
for $N= N_1 p + N_0$.
Here, whether $\nu_j$ is zero or not is finally determined by the choice of the boundary transfer matrix $T_0$.
Since one of principal eigenvalues of $T$ is real,
which will be shown in our numerical calculations,
we will choose $T_0$ to eliminate contribution from the non-real principal eigenvalue
and to set $\nu_j=\delta_{j1}$ supposing $\theta_1 =0$,
which means the $p$-site translational invariance
and enables us to express the expectation values in a simple form.

\section{Calculation of Expectation Values}
For the calculation of the bulk energy 
$\varepsilon=\lim_{N\rightarrow\infty} {\bra{\Psi}{\cal H}\ket{\Psi} \over N \braket{\Psi}{\Psi}}$,
we can neglect the boundary effect which gives the ${\cal O}(1/N)$ boundary energy.
Here the local Hamiltonian ${\cal H}_{[i,i']}$ are supposed to include $i'-i+1$ spins from $i$th site to $i'$th site.
Then, after using $\nu_j=\delta_{j1}$,
we obtain
\begin{math}
  \varepsilon 
  = 
  {\vec{v}_1^\dagger h
    \vec{u}_1 
 }
\end{math}
with
\begin{eqnarray*}
  h_{\alpha\alpha''\sigma_1;\alpha'\alpha'''\sigma_{N_h p}}
  &=&  
  {\bra{M^{\sigma_1\sigma_{N_h p}}_{\alpha\alpha'}}
  {\cal H}_{[2,N_h p-1]}
  \ket{M^{\sigma_1\sigma_{N_h p}}_{\alpha''\alpha'''}}
  \over 
  N_h p-2}
  ,
\end{eqnarray*}
where
$\ket{M^{\sigma_1\sigma_{N_h p}}_{\alpha\alpha'}}=
\sum_{\bfsigma}
\ket{\bfsigma}
\left(\prod_{i=1}^{N_h p-1} A_{i}^{\sigma_{i}\sigma_{i+1}}\right)_{\alpha\alpha'}$
with $\bfsigma=\sigma_{2},\ldots,\sigma_{N_h p-1}$.
Here $N_h$ is supposed to be enough large to include the periodicity of the Hamiltonian.
In the same manner, one can calculate any expectation values per site.
To optimize the IRF-type MPS the downhill simplex method and the modified Powell method are used.~\cite{Numerical_Recipes:book}

\section{Principal Eigenvalues and Spontaneous Symmetry Breaking}
To reveal the relation between principal eigenvalues $\lambda_j$ with $p=1$ and the spontaneous symmetry breaking of the translational symmetry, first, we discuss the magnetization plateau state of the classical $S=1/2$ Ising model
with the third next nearest neighbor interaction given by 
\begin{math}
{\textstyle {\cal H}_{\rm I}(t) = \sum^{}_{i} \left( H_{}(t) S^z_i +  \sum^{3}_{k=1} J^{z}_{k}(t) S^z_i S^z_{i+k} \right) }, 
\end{math}
where $S^z_i$ represents the $z$ component of the spin-$1/2$ operator. 
The longitudinal magnetic field and $k$-th nearest neighbor interaction are given by $H_{}$ and $J^{z}_{k}$ respectively.
To show various magnetic phases of the ground state, all parameters are defined as a function of one parameter $t\in [0,1]$ as follows;
\begin{math}
H_{}(t) = 0.9 t, 
J^{z}_{1}(t) = -0.14	t + 0.57 ,
J^{z}_{2}(t) = -0.22	t + 0.46 ,
J^{z}_{3}(t) = -0.48 t + 0.38.
\end{math}
In fact, we detect six different phases in the parameter region of $t$ with using analytic results.

Figure 2 shows the magnetization per site as a function of $t$ obtained by the IRF-type MPS with $p=1$ and $m=3$,
which completely agrees with that of the analytic results. 
The spin configuration from analytic results in each phase are also denoted by the black allow. 
The six diagrams mean the complex principal eigenvalues $\lambda_j$ of the matrix $T$ in each phase. 
It is shown that $q$ principal eigenvalues correspond to
the spontaneous symmetry breaking of the $q$-site translation.
If we enlarge the periodicity $p$ to $q$,
analytic results can be exactly represented at $m=1$.
In this case, the unique principal eigenvalue is obtained. 
\begin{figure}
  \centering
  \resizebox{7cm}{!}{\includegraphics{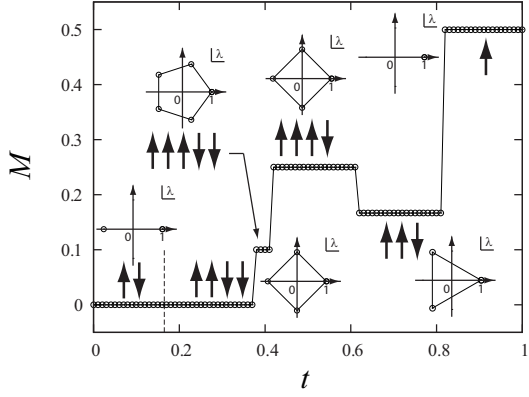}}\\
  \caption{Magnetization per site $M$ and principal eigenvalues of the matrix $T$ as a function of the model parameter $t$ in the Hamiltonian ${\cal H}^{}_{{\rm I}}(t)$.}
  \label{fig:2}
\end{figure}

\section{Result:Heisenberg chain}
Secondly, we show the result of the uniform quantum chain. 
The Hamiltonian of the Heisenberg chain with the longitudinal magnetic field is given by 
\begin{math}
{\cal H}_{\rm HB} = {\textstyle \sum^{}_{i} \left( H_{} S^z_i +  J {\bf S}^{}_i \cdot {\bf S}^{}_{i+1} \right)},  
\end{math}
where ${\bf S}^{}_i$ represents the spin-$S$ operator. 
Here, we take $S=1/2$ and $1$. 
The parameter $J$ means the intensity of the isotropic exchange interaction, and it is taken as the unit of energy. 
The MPS with the period $p=2$ is selected in both spin-$S$ chains, 
because the MPS with $p=2$ gives the minimum energy for fixed $m$ within $p\leq 6$,
which reminds us N\'{e}el state's periodicity.

Figure 3 shows the magnetization curve for (a) $S=1/2$ and (b) $S=1$ Heisenberg chains. 
The dimension of the matrix $m$ is up to three in each chain. 
As reference data, we also show (a) the rigorous result in $S=1/2$ from the Bethe ansatz, 
and (b) the reported data from the DMRG calculation,~\cite{Nishino:Butsuri55} respectively. 
The results of our MPS of $m=3$ agree with the references in this scale. 
\begin{figure}
  \centering
  \resizebox{7cm}{!}{\includegraphics{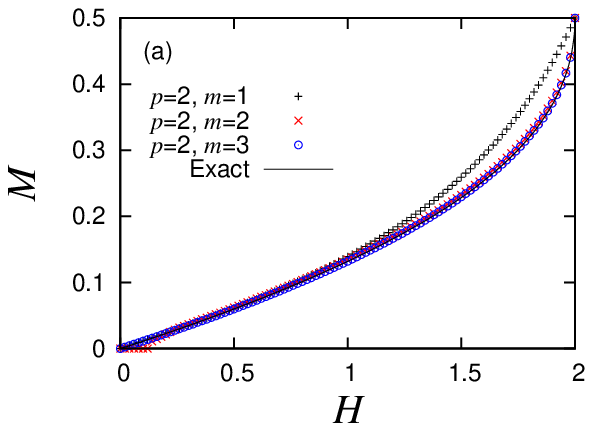}}\\
  \resizebox{7cm}{!}{\includegraphics{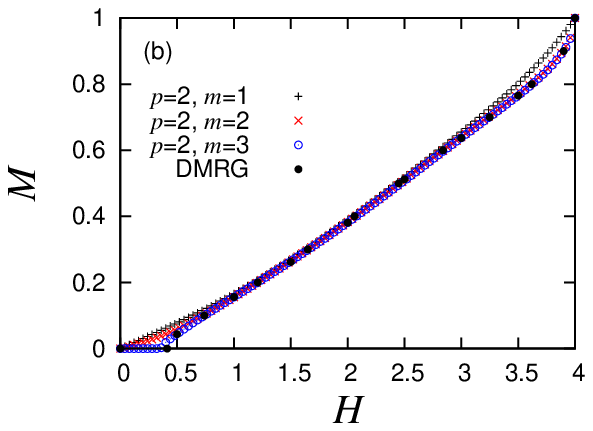}}\\
  \caption{Magnetization per site $M$ as a function of magnetic field $H_{}$
    for (a)$S=1/2$ and (b)$S=1$ Heisenberg chains
    with the periodicity $p$ and dimension $m$.
  }
  \label{fig:3}
\end{figure}

\section{Result: Okunishi}
Finally, we show the result of $S=1/2$ Ising-like zigzag $XXZ$ chain, which is a uniform quantum frustrated chain. 
The Hamiltonian is given by 
\begin{math}
{\cal H}_{\rm XXZ}  = 
{\textstyle \sum^{}_{i}} ( H_{} S^z_i + 
 {\textstyle \sum_{k=1}^{2}} J^{z}_{k}( \Delta ( {S}^{x}_i {S}^{x}_{i+k} + {S}^{y}_i {S}^{y}_{i+k} ) + {S}^{z}_i {S}^{z}_{i+k} ) ), 
\end{math}
with the anisotropic parameter $\Delta$. 
We choose parameters as $\Delta = 0.1$, $J^{z}_{1}=0.4$, and $J^{z}_{2}=1.0$. 
It is known that the 1/3-plateau state appears in this system under the parameters.~\cite{Okunishi:PRB68} 
In addition, $q=3$ principal eigenvalues are confirmed in the matrix $T$ with the period $p=1$
in the 1/3-plateau.
The DMRG data shown by the black line in Fig. 4 are also re-calculated following Ref. \citen{Okunishi:PRB68}).
The perfect 1/3 plateau is obtained by the MPS in small $m$, while there is the finite-size effect for the DMRG result.
In the plateau we obtain the minimum energy at $p=3$.
Unlike ${\cal H}_{\rm HB}$, ${\cal H}_{\rm XXZ}$ shows the transition from $p=3$ to $p=4$ in large magnetic fields as shown in Fig. 4. 
The transition is artificial due to small $m$, but the DMRG data also show the property change about the stability of the odd $M$ states.
It can be concluded that more large $m$ is needed. 
\begin{figure}
  \centering
  \resizebox{8cm}{!}{\includegraphics{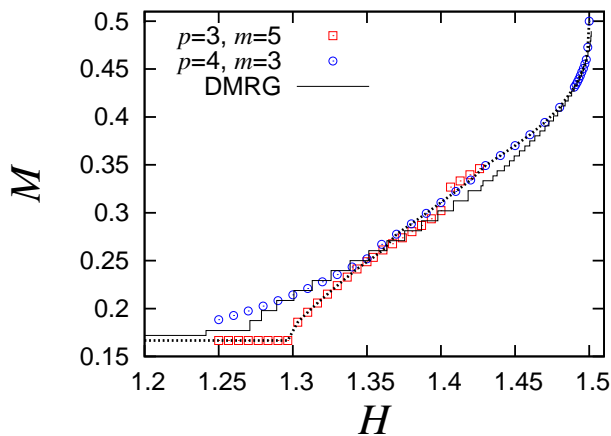}}\\
  \caption{Magnetization per site $M$ as a function of magnetic field $H$
    for the $S=1/2$ zigzag chain $\mathcal{H}_{\rm XXZ}$. 
    The DMRG data is calculated for 192 sites with the open boundary condition. 
    The dotted line means the magnetization $M$ of the best variational state. 
  }
  \label{fig:4}
\end{figure}

\section{conclusion}
In summary,
we investigated the uniform IRF-type MPS with an arbitrary period $p$ and with a boundary matrix as a generalization of the usual MPS.
The boundary matrix related to the periodicity is used to hold the $p$-site translational invariance in the numerical calculation for the classical or quantum spin chain in the thermodynamic limit.
We found that the number of the principal eigenvalues, $q$, in the matrix $T$ with $p=1$ is consistent with the period of the spontaneous symmetry braking ground state in the classical Ising model and indicates the optimum periodicity which gives the minimum energy for the fixed dimension of the artificial basis $\alpha$.
The IRF-type MPS with the optimum periodicity $p=2$ gives consistent magnetization curves in $S=1/2$ and $1$ Heisenberg chains and that with $p=3$ captures the correct magnetization at the 1/3 plateau without the finite-size effect in the $S=1/2$ zigzag chain.
This means that the plateau state is easy to reproduce via this variational state which has small quantum entanglement except for around the upper critical magnetic field of the 1/3 plateau.
Our demonstration is limited for small $m$, but smallness of $m$ leads us to the optimum periodicity $p=4$ shown in the magnetization curve in zigzag chain, which can be related to property of the true ground state.
To establish an optimization method for large $m$ is one of future works.

Note that the period of the MPS is important for the wave function prediction in the infinite algorithm of the DMRG method efficiently.~\cite{JPSJ.64.4084, JPSJ.75.014003, JPSJ.77.114002, 0804.2509, JPSJ.79.044001} 
The number of principal eigenvalues in the matrix $T$ is also meaningful information for the prediction. 

%%% Local Variables:
%%% TeX-master: "jpsj"
%%% End:

\acknowledgement 

This work was supported in part by Grant-in-Aid for JSPS Fellows, Grant-in-Aids (No. 20740214), Global COE Program (Core Research and Engineering of Advanced Materials-Interdisciplinary Education Center for Materials Science) from the Ministry of Education, Culture, Sports, Science and Technology of Japan. 

\bibliography{my_list.bib,comment.bib}

\begin{thebibliography}{10}
\bibitem{comment:MPS1}
{If we suppose $A_i^\sigma$ are matrices,we can find the representation:
  \begin{math}{\tiny \tilde{A}_{1} = \left(
  \begin{array}{cc}-2^{-1/4}\ket{\uparrow} & 0 \\2^{-1/4}\ket{\downarrow} & 0
  \\\end{array}\right),~\tilde{A}_{2} = \left(
  \begin{array}{cc}2^{-1/4}\ket{\downarrow} & 2^{-1/4}\ket{\uparrow} \\0 & 0
  \\\end{array}\right) } \end{math}}.
\bibitem{comment:MPS2}
{$A^{\uparrow \uparrow}_{1}=A^{\downarrow \downarrow}_{1}=0$, $A^{\uparrow
  \downarrow}_{1}=-A^{\downarrow \uparrow}_{1}=1/\sqrt{2}$, and $A^{\uparrow
  \uparrow}_{2}=A^{\downarrow \downarrow}_{2}=A^{\downarrow
  \uparrow}_{2}=A^{\uparrow \downarrow}_{2}=1$, where the dimer is on
  $(2i-1)$--$(2i)$ sites}.
\end{thebibliography}
%\bibliography{macro,wiki,book,mybib,my_list.bib}
%\input{jpsj.bbl.save}

%\newpage
%\input{noteall}

\end{document}